\pdfoutput=1
\documentclass[
	physrev,
	reprint,
	amsmath,
	amsfonts,
	]{revtex4-2}

\usepackage[all]{nowidow}
\usepackage{csquotes}
\usepackage{float}

\usepackage{mathtools}

\usepackage[detect-all]{siunitx}
\usepackage{dcolumn}
\newcommand{\EQ}[3]{
  \begin{equation}
    \label{#1}
    #2
    \;#3
  \end{equation}
}

\def\hat{\mathaccent "705E\relax}

\begin{document}

\title{Time-dependent condensate formation in ultracold atoms with energy-dependent transport coefficients}

\author{M. Larsson}
%\email{hoelck@thphys.uni-heidelberg.de}
\author{G. Wolschin}
\email{wolschin@thphys.uni-heidelberg.de}
\affiliation{Institute for Theoretical Physics, Heidelberg University, Philosophenweg 16, 69120 Heidelberg, Germany, EU}

%\day=14
%\month=9
%\year=2020
\date{\today}

\begin{abstract}
Time-dependent Bose--Einstein condensate (BEC) formation in ultracold atoms is investigated in a nonlinear diffusion model. For constant transport coefficients, the model has been solved analytically. Here, we extend it to include energy-dependent transport coeffcients and numerically solve the nonlinear equation. Our results are compared with the earlier analytical model for constant transport coeffcients, and with recent deep-quench data for $^{39}$K at various scattering lengths. Some non-physical predictions from the constant-coefficient model are resolved using energy-dependent drift and diffusion.
%We use the results to estimate a condensate-formation timescale for ultracold fermionic $^6$Li and bosonic $^7$Li systems. 
\end{abstract}

\maketitle

\section{Introduction}
\label{intro}
Time-dependent \textcolor{black}{Bose--Einstein condensate formation} in ultracold quantum gases is an ideal testing ground for nonequilibrium-statistical models and theories. Following the early 1995 BEC discoveries in alkali atoms, the time-dependence of condensate formation in the course of evaporative cooling was first studied 1998 in $^{23}$Na \cite{miesner1998} and compared with available numerical model results \cite{bzs00}. Here, a short radio-frequency pulse was used to remove high-energy atoms in order to investigate the subsequent time-dependent thermalization and transition to the condensed state.

 Much more precise experimental data with error bars were obtained 2002 for $^{87}$Rb \cite{kdg02}, where cooling into the quantum-degenerate regime was achieved by continuous evaporation for a duration of several seconds. The data were compared with the results of a numerical model based on quantum kinetic theory \cite{gz97} that had been developed at a time when no data were available, predicting a thermalization timescale for  $^{87}$Rb of the order of 5 seconds, more than one order of magnitude larger than the one for $^{23}$Na. The model was later adapted to the data  \cite{kdg02} when the measured thermalization timescale in rubidium turned out to be similar to the one in sodium.
 
For a detailed comparison of nonequilibrium-statistical models for time-dependent BEC formation and data, it appears more promising to start from a thermalized atomic cloud in a trap that is subjected to a short deep quench, because the initial condition is fixed and not time-varying as in continuous evaporative cooling. Corresponding time-resolved data were provided by the Cambridge group in 2021 using a homogeneous 3D Bose gas of $^{39}$K with tunable interactions and near-perfect isolation in a cylindrical optical box \cite{gli21}. We have compared these data with results of our nonlinear diffusion model \cite{gw18,gw22} in Ref.\, \cite{kgw22}. At four different interaction strengths, we found agreement with the data already for constant transport coefficients, where the model is analytically solvable for deep-quench initial conditions given by a truncated Bose--Einstein distribution, and boundary conditions at the singularity \cite{rgw20}.

 In the present work, we extend this approach to energy-dependent transport coefficients. This case is physically more realistic, and artifacts of the analytical solution can be resolved. On the downside, the nonlinear boson diffusion equation (NBDE) is no longer solvable with analytical methods. However, in the limit of constant coefficients, the precision of our numerical solution can be checked through a comparison with the exact result, which is rarely possible in other kinetic approaches, with the exception of a linear relaxation-time approximation to the nonlinear thermalization problem.

In the following section, we briefly review the nonlinear boson diffusion model, and its analytical solution for constant coefficients. Energy-dependent drift and diffusion coefficients are introduced in Sect.\,III, the numerical solution of the nonlinear equation is presented, and in the special case of constant transport coefficients compared with the available exact solution. The time-dependent condensate fraction for  $^{39}$K following a deep quench that removes 77\% of the atoms and $\simeq97.5\%$ of the energy is calculated numerically for various scattering lengths in Sect.\,IV, and compared to the Cambridge data. 
%We also estimate the condensate-formation timescale for ultracold fermionic $^6$Li and bosonic $^7$Li systems. 
The conclusions are drawn in Sect.\,V.

%\cite{gli21,gz97,rgw20,sgw21}

\section{Nonlinear diffusion model}
\label{model}
With our previously developed \cite{gw18,gw22} nonlinear model we have already computed the nonequilibrium evolution and time-dependent condensate formation in equilibrating Bose gases of ultracold atoms using constant transport coefficients for $^{23}$Na \cite{sgw21}, and more recently \cite{kgw22} with a comparison to deep-quench data \cite{gli21} for $^{39}$K.
We describe the cold-atom vapor in a trap as a time-dependent mean field with a collision term and consider only $s$-wave scattering 
with the corresponding scattering lengths. The $N$-body density operator ${\hat{\rho}_N(t)}$ is composed of $N$ single-particle wave functions of the atoms which are solutions of the time-dependent Hartree-Fock equations plus a time-irreversible collision term $\hat{K}_{N}(t)$ . The latter causes the system to thermalize through random two-body collisions 
($\hbar=c=1$)
\EQ{}{
i\,\frac{ \partial\hat{\rho}_N(t)}{\partial t} = \big[\hat{H}_\mathrm{HF}(t),\hat{\rho}_{N}(t)\big] + i \hat{K}_{N}(t)\,.
\label{eq1}}{}
The self-consistent Hartree-Fock mean field of the atoms is $\hat{H}_\mathrm{HF}(t)$, where the trap provides an external potential.

The full many-body problem can be reduced to the one-body level in an approximate version for the ensemble-averaged single-particle density operator 
 $\bar{\rho}_1(t)$  {by taking the trace over particles $2...A$}. Its diagonal elements represent the 
probability for a particle to be in a state $|\alpha\rangle$ with energy $\epsilon_\alpha$
\EQ{}
{\big(\bar{\rho}_1(t)\big)_{\alpha,\alpha} = n(\epsilon_\alpha, t)\equiv n_\alpha(\epsilon,t)}
{.}
The total number of particles is $N=\sum_\alpha n_\alpha$. We neglect
the off-diagonal terms of the density matrix, {because their time evolution mainly consists of a loss term that tends to dissipate away the non-diagonal elements generated by the time evolution of the mean field, as was shown for fermions \cite{gww81}, and is expected to occur similarly for bosons}.

{The many-body Hartree-Fock term is reduced to the one-body level as well, with the mean-field Hamiltonian plus the confining potential that represents the trap,
but in the following we concentrate on the collision term that causes the thermalization of the system, and also drives the time-dependent formation of the Bose--Einstein condensate.}

Assuming \textit{sufficient ergodicity} as has been widely discussed in the literature  \cite{snowo89,kss92,setk95,lrw96,hwc97,jgz97}, the occupation-number distribution $n_\alpha(\epsilon,t)$ in a finite Bose system obeys a Boltzmann-like collision term, where the distribution function depends only on energy and time.
{For elastic two-body interactions, the equation for the
single-particle occupation numbers $n_1(\epsilon,t)$ can be expressed as \cite{gw18}
\begin{eqnarray}
\frac{\partial n_1}{\partial t}=\sum_{\epsilon_2,\epsilon_3,\epsilon_4}^\infty  \langle V_{1234}^2\rangle\,G\,(\epsilon_1+\epsilon_2, \epsilon_3+\epsilon_4)\times\qquad\qquad\\ \nonumber
\bigl[(1+n_1)(1+n_2)\,n_3\,n_4-
(1+n_3)(1+n_4)\,n_1\,n_2\bigr]
 \label{boltzmann}
\end{eqnarray}
with the second moment of the interaction $\langle V^2\rangle$ and the function $G$, which ensures energy conservation. 
In an infinite system, it would be a $\delta$-function  as in the usual Boltzmann collision term
where the single-particle energies are time independent,
\begin{equation}
G\,(\epsilon_1+\epsilon_2, \epsilon_3+\epsilon_4)\rightarrow \pi\, \delta(\epsilon_1+\epsilon_2- \epsilon_3-\epsilon_4)\,.
 \label{delta}
\end{equation}
In a finite system, however, the energy-conserving function acquires a width,
such that off-shell scatterings between single-particle states which lie apart in energy space become possible \cite{gww81}. }

The reduction to $1+1$ dimensions corresponds to spatial and momentum isotropy. For the thermal cloud of cold atoms surrounding a Bose--Einstein condensate (BEC), this is expected to be approximately fulfilled. Different spatial dimensions enter our present formulation through the density of states, which depends also on the type of confinement. The model calculations in this work are for a 3D system, and as in the $^{39}$K experiment \cite{gli21}, we investigate results for the density of states of bosonic atoms confined in a cylindrical optical box.

{From the above quantum Boltzmann collision term, we had derived in \cite{gw18,gw22}
the nonlinear boson diffusion equation (NBDE)
 for the single-particle occupation probability distributions $n\equiv n\,(\epsilon,t)$ as}
 \begin{equation}
	%\frac{\partial n}{\partial t}=-\frac{\partial}{\partial{\epsilon}}\left[\varv\, n\,(1\pm n)+n\frac{\partial D}{\partial \epsilon}\right]+\frac{\partial^2}{\partial{\epsilon}^2}\bigl [D\,n\,\bigr]
	{\partial_t n}=-{\partial_{\epsilon}}\left[v\, n\,(1+ n)+n\,{\partial_\epsilon D}\right]+{\partial_{\epsilon\epsilon}}\bigl [D\,n\,\bigr]\,,
\label{nbde}
\end{equation}
%where the $+$ sign represents bosons, and the $-$ sign fermions. 
The drift term $v\,(\epsilon,t)<0$ accounts for the shift of the distribution towards the infrared, the diffusion function $D\,(\epsilon,t)$ for the broadening with increasing time. {Both transport coefficients are defined in terms of moments of the transition probabilities.} In the present case of a deep quench in a thermalized vapor, the diffusion coefficient causes the softening of the sharp cut at  $\epsilon=\epsilon_\mathrm{i}$ in the UV that signifies the quench, as well as the diffusion of particles into the condensed state in the IR.
The many-body physics is contained in these
transport coefficients, which depend on energy, time, and the second moment of the interaction. 
 The derivative term of the diffusion coefficient in the NBDE (\ref{nbde}) causes
% as compared to ref.\,\cite{gw18} 
 the stationary solution $n_\infty(\epsilon)$ to become a Bose--Einstein distribution, which is attained for $t\rightarrow\infty$ as
\begin{equation}
n_\infty(\epsilon)=n_\mathrm{eq}(\epsilon)=\frac{1}{e^{(\epsilon-\mu)/T}- 1}\,,
 \label{Bose--Einstein}
\end{equation}
under the condition that the ratio $v/D$ has no energy dependence, and requiring that
%for the limit of time to infinity
$\lim_{t\rightarrow \infty}[-v\,(\epsilon,t)/D\,(\epsilon,t)] \equiv 1/T$. (We use units $k_\text{B}=\hbar=c=1$ throughout this manuscript).
The chemical potential is $\mu\leq0$ in a finite Bose system. It appears as an integration constant in the stationary solution of 
Eq.\,(\ref{nbde}) for $t\rightarrow\infty$. It vanishes in case of inelastic collisions that do not conserve particle number such that the temperature alone determines the equilibrium distribution. 

In the limit of energy-independent transport coefficients $D, v$, the fluctuation-dissipation relation is simply $T=-D/v$, and the NBDE can be solved analytically through a nonlinear transformation \cite{gw18,gw22}. It is one one the few nonlinear partial differential equations that have a clear physical meaning and can be solved exactly. The solution is
\begin{equation}
	n\,(\epsilon,t) =T {\partial_{\epsilon}}\ln{\mathcal{Z}(\epsilon,t)} -\frac{1}{2}= \frac{T}{\mathcal{Z}} {\partial_\epsilon\mathcal{Z}} -\frac{1}{2}
	\label{eq:Nformula}	
\end{equation}
with the time-dependent partition function ${\mathcal{Z}(\epsilon,t)}$ that fulfills the linear diffusion (heat) equation 
  \begin{equation}
\partial_t\,\mathcal{Z}(\epsilon,t)=D\, {\partial_{\epsilon\epsilon}}{\,\mathcal{Z}(\epsilon,t)}\,.
    \label{eq:diffusionequation}
\end{equation}
The time-dependent partition function $\mathcal{Z}(\epsilon,t)$ and its energy-derivative are given analytically, but still require integrals over the bounded Green's function of the linear problem
and an exponential function that contains an energy-integral over the initial distribution. However, for certain simplified initial conditions such a theta-function \cite{gw18}, or a truncated Bose--Einstein distribution \cite{rgw20}
\begin{equation}
n_\mathrm{i}(\epsilon)=\frac{1}{\exp{(\frac{\epsilon-\mu_\mathrm{i}}{T_\mathrm{i}})}-1}\theta(1-\epsilon/\epsilon_\mathrm{i})\,,
\label{ni0}
\end{equation}
exact analytical results have been obtained and used in comparisons of the time-dependent condensate fraction to data \cite{sgw21,kgw22}. For the present case of a Bose--Einstein distribution
at temperature $T_\text{i}$ truncated via a deep quench at an energy $\epsilon_\text{i}$ such that the temperature of the final equilibrium distribution becomes $T_\text{f}=-D/v$, the generalized partition function can be expressed as an infinite series \cite{rgw20}
\begin{equation}
{\mathcal{Z}}(\epsilon,t) = \sqrt{4 D t} \, \exp\Bigl(-\frac{\mu}{2 T_{\mathrm{f}}}\Bigr) \sum_{k=0}^{\infty} \binom{\frac{T_{\mathrm{i}}}{T_{\mathrm{f}}}}{k} \left( -1 \right)^k \times
f_k^{\text{T}_\text{i},\text{T}_\text{f}} \,(\epsilon,t)\,,
\label{partfct}
\end{equation}
and similarly for the energy-derivative of $Z(\epsilon,t)$.
The exact analytical expressions for $f_k^{\text{T}_\text{i},\text{T}_\text{f}} \,(\epsilon,t)$ are combinations of exponentials and error functions. The distribution function can now be computed directly from Eq.\,(\ref{eq:Nformula}). The convergence of the solutions has been studied in  Ref.\,\cite{sgw21}:
a result with $k_\mathrm{max}=10-40$ expansion coefficients -- depending on the system and the observable --  is indistinguishable from the exact solution with $k_\mathrm{max}=\infty$. The infinite series terminates in case $T_\mathrm{i}/T_\mathrm{f}$ is an integer due to the binomial coefficient, and we explicitly use that property for $^{39}$K, where the exact solution are obtained for
$k_\mathrm{max}=4$. A brief summary of the nonlinear diffusion model with constant transport coefficients for both, boson and fermion systems at low and high energies and temperatures is presented in Ref.\,\cite{gw22}.

\section{NBDE with energy-dependent drift and diffusion}
Whereas the exact solution for constant transport coefficients already yields physically reasonable results, it also has drawbacks. For example, when calculating the time-dependent condensate fraction using the additional condition of particle-number conservation, it may slightly overshoot the equilibrium value and approach it from above at later times:
A large constant drift towards the IR causes thermalization in the UV region to be too slow, such that at large times the Bose--Einstein limit is reached in the IR, but not yet in the UV, resulting in an unbalanced energy-dependent equilibration. Although this effect is often hardly visible in a plot, it is a deficiency of the constant-coefficient case that can be remedied using energy-dependent drift and diffusion coefficients.

For energy-dependent transport coefficients, we presently cannot solve the NBDE (\ref{nbde}) analytically, although for simplified cases such as constant diffusion and linear drift one may try to find solutions. We shall therefore resort to numerical solutions using the Julia programming language. In the limiting case of constant transport coefficients, we shall compare the numerical solutions to the exact results in order to assess the accuracy of the numerical approach.
\subsection{Energy dependence of the transport coefficients}
We seek energy-dependent transport coefficients that cause a faster thermalization in the region $\epsilon > \epsilon_\text i$ and yield the correct final distribution of particles in the limit $t \rightarrow \infty$. In a semi-classical description, the diffusion at $\epsilon = 0$ is zero such that particles with zero momentum cannot contribute to the transport of particles into non-zero energy states. However, in a quantum description, the transport coefficients at $\epsilon = 0$ can take a finite value, denoted as $v_0$ and $D_0$. For a system of bosons, which has a singularity at the chemical potential, a time-independent boundary condition can be assigned at $\epsilon = \mu = 0$. The position of the singularity is then fixed, the one-particle distribution function does not change with time at $\epsilon = 0$, and the constraint $v_0 = D_0 = 0$ at $\epsilon = 0$ applies. For an initial distribution without a singularity at the chemical potential, like a box distribution for a gluon system at relativistic energies, a time-dependent boundary at $\epsilon = \mu$ is, however, required.
      
The solution of the nonlinear boson diffusion equation should still converge to the Bose--Einstein equilibrium distribution for $t\rightarrow \infty$. 
%By changing this constraint and therefore the equilibrium solution of the system, one could account for changes in the condensate fraction due to corrections in the critical temperature %for example. However, since the previously outlined theory only includes the critical temperature of an ideal Bose gas and thus the corresponding equilibrium condensate fraction, we %stick to the ideal Bose gas properties for now and demand the transport coefficients to yield the expected Bose--Einstein distribution in the stationary limit, $n_{\infty} = [\text{exp}
%(\epsilon/T)-1]^{-1}$. 
Therefore, when substituting $n(\epsilon, t)$ with $n_\infty$ in Eq.\,(\ref{nbde}), the following stationary nonlinear diffusion equation must hold
\begin{align}
0 &= -\partial_{\epsilon}\Big[v(\epsilon)n_{\infty}\bigl(1+n_{\infty}\bigr)\Big] + \partial_{\epsilon}\Big[D(\epsilon)\partial_{\epsilon}n_{\infty}\Big] \\
&= \frac{\partial v}{\partial \epsilon} + v \cdot \underbrace{\frac{\partial_{\epsilon}\big[n_{\infty}\bigl(1+n_{\infty}\bigr)\big]}{n_{\infty}\bigl(1+n_{\infty}\bigr)}}_{\equiv p(\epsilon)} -\underbrace{\frac{\partial_{\epsilon}\big[D(\epsilon)\partial_{\epsilon}n_{\infty}\big]}{n_{\infty}\bigl(1+n_{\infty}\bigr)}}_{\equiv q(\epsilon)} \, ,
\end{align}
which is an inhomogeneous partial differential equation for $v(\epsilon)$ depending on a unknown function $D(\epsilon)$ and has the general solution
\begin{equation}
v(\epsilon) = \frac{1}{\text{exp}\bigl(\int_0^{\epsilon}p(\epsilon)\;d\epsilon\bigr)}\Bigg[\int_0^{\epsilon} \text{exp}\biggl(\int_0^{\Tilde{\epsilon}} p(\Tilde{\epsilon}) \;d\Tilde{\epsilon}\biggr)q(\epsilon)\;d\epsilon + v_0 \Bigg] \, .
\label{gs}
\end{equation}
The integral over $p(\epsilon)$ is
\begin{equation}
\int_0^{\epsilon}p(\epsilon)\;d\epsilon = \int_0^{\epsilon} \partial_{\epsilon}\, \text{ln}\Big[n_{\infty}\bigl(1+n_{\infty}\bigr)\Big]\;d\epsilon = \text{ln}\Big[n_{\infty}\bigl(1+n_{\infty}\bigr)\Big] \, ,
\end{equation}
{such that the integrals inside the square brackets in Eq.\,(\ref{gs})} simplify to
\begin{align}
&\int_0^{\epsilon} \frac{n_{\infty}(1+n_{\infty})}{n_{\infty}(1+n_{\infty})}\biggl\{ -\partial_{\epsilon}\big[D(\epsilon)\partial_{\epsilon}n_{\infty}\big] \biggr\} d\epsilon =- D(\epsilon) \partial_{\epsilon}n_{\infty} \\
&\Rightarrow v(\epsilon) =- \frac{D(\epsilon)\partial_{\epsilon}n_{\infty} + v_0}{n_{\infty}(1+n_{\infty})} \, .
\end{align}
Applying now the constraint $v(\epsilon=0) = v_0 \equiv 0$ and inserting the Bose--Einstein distribution for $n_{\infty}$ yields the previously discussed relation between the transport coefficients
\begin{equation}\label{eq:25}
-\frac{D(\epsilon)}{v(\epsilon)} = \frac{1 + n_{\infty}}{n_{\infty}e^{(\epsilon/T_\text{f})}}\, T_\text{f}= T_\text{f} \, .
\end{equation}
Hence, for each time step at any energy, the relation has to be fulfilled to reach the desired Bose--Einstein distribution in the stationary limit. 

\begin{figure}
%\resizebox{0.48\textwidth}{!}%
  \includegraphics[scale=0.56]{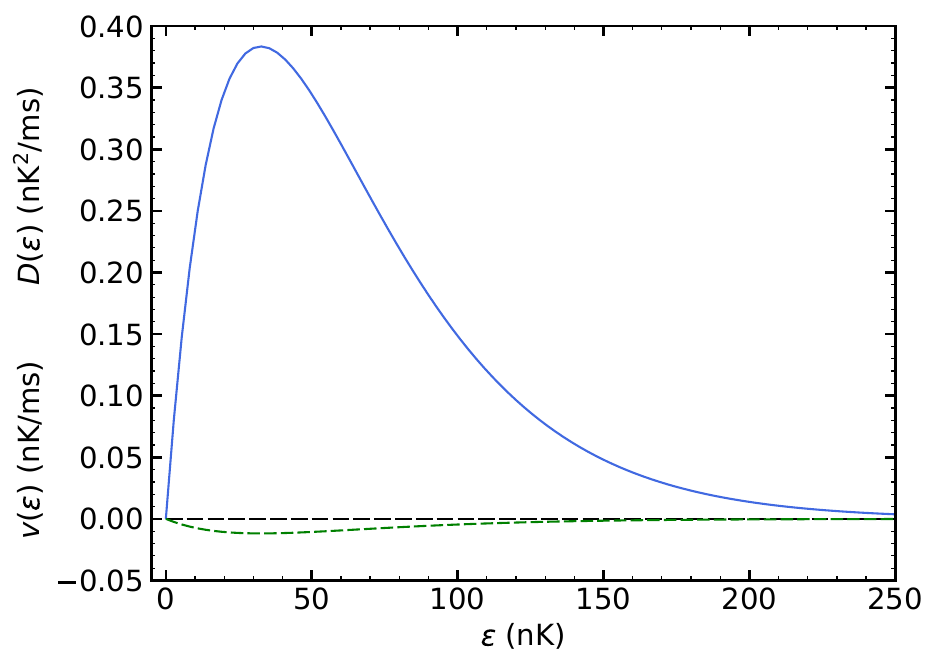}
\caption{Energy-dependent transport coefficients for an equilibrating 3D Bose gas of $^{39}$K in an optical box for a single energy quench from $T_\text i = 130\,\text{nK}$ to $T_\text f = 32.5\,\text{nK}$ with $\mu_\text i = -0.67\,\text{nK}$ and $\epsilon_\text i = 15.56\,\text{nK}$, calculated from a Boltzmann-distribution ansatz. The plot shows the diffusion coefficient $ D(\epsilon)$ in $(\text{nK}^2/\text{ms})$ ({upper solid curve}) and the drift term $ v(\epsilon)$ in $(\text{nK}/\text{ms})$ ({lower dashed curve}) {for $a=140\,a_0$}.
}
\label{fig1}       % Give a unique label
\end{figure}

 To obtain a system that thermalizes into a condensed and non-condensed phase, one has to undercut the critical temperature $T_\text c$. Experimentally this is done by evaporative cooling or by sudden energy quenches, such that all particles with an energy $\epsilon > \epsilon_\text i$ are removed from the trap. Therefore one can steer the final temperature by changing the quench energy and one gets $T_\text f < T_\text c < T_\text i$, where $T_\text i$ is the initial temperature before the quench and $T_\text f$ is the equilibrium temperature  that is eventually reached after removing higher-energy particles from the trap. In \cite{sgw21} it is shown that no condensate forms if the final temperature $T_\text f$ at the critical number density $n_\text c$ stays above the critical temperature $T_\text c$ for bosonic atoms of mass $m$,
\begin{equation}
    T_\text c = \frac{2\pi}{m}\biggl(\frac{n_\text c}{\zeta(\frac{3}{2})}\biggr)^{2/3} \, ,
\end{equation}
where we neglect the effects of interactions \cite{gro95} at the moment.
The order of $T_\text c$ for a non-relativistic
3D gas inside a box trap is $\mathcal{O}(100\, \text{nK})$ and hence the final temperatures of experimentally reasonable setups range between $10\,\text{nK}$ and $500\,\text{nK}$ for common BEC experiments with $^{87}$Rb, $^{39}$K, $^{23}$Na and $^7$Li of approximately $10^5$ particles inside a box trap.\\
In the simplified model with constant transport coefficients, the values of $v$ and $D$ are connected to the equilibrium temperature $T_\text f = -D/v$ and the local thermalization time $\tau_\text{eq}$. This relationship arises from the equality of the stationary solution of the nonlinear diffusion equation with the Bose–Einstein distribution. The thermalization time at the cut $\epsilon = \epsilon_\text i$ has been deduced from an asymptotic expansion of error functions in the analytical solutions of the non-linear diffusion equation for a theta-function initial condition, as $\tau_\text{eq} \equiv \tau_{\infty}(\epsilon = \epsilon_\text i) \simeq 4D/(9v^2)$ \cite{gw18}. For a general initial distribution with a cut at $\epsilon = \epsilon_\text i$, it can be expressed as
\begin{equation}
    \tau_\text{eq} \equiv \tau_{\infty}(\epsilon = \epsilon_\text i) \simeq f\,D/v^2\, ,
\end{equation}
where $f = 4$ for fermions and $f = 4/9$ for bosons in the case of an initial theta-function distribution. The analytical result for a truncated Bose–Einstein initial distribution has not been derived yet but is expected to be considerably shorter than that for a theta-function distribution because the shape of the initial distribution resembles already the final Bose–Einstein result. For $^{39}$K,  $f = 0.045$ was adopted in \cite{kgw22} to calculate the transport coefficients such that they align with the thermalization time of experimental data. With $f$ being a required experimental constant, the transport coefficients for the analytical solution of the nonlinear Boltzmann diffusion equation are thus determined by
\begin{align}
    D &= \frac{f \; T_\text f^2}{\tau_\text{eq}} \, ,  \label{eq:26}\\
    v &= -\frac{f \; T_\text f}{\tau_\text{eq}} \, . \label{eq:27}
\end{align}
\begin{table}
%Check -0.00246 or -0.00256!
\begin{center}
\caption{Average values of the transport coefficients, initiation and equilibration times for BEC formation in $^{39}$K}($T_\mathrm{i}=130$ nK, $T_\mathrm{f}=-\langle D \rangle/\langle v\rangle=32.5$ nK)
\vspace{.2cm}
\label{tab1}       % Give a unique label
% For LaTeX tables use
\begin{tabular}{rrrrr}
\hline\noalign{\smallskip}
$a\,(a_0)$&$\langle D \rangle$\,(nK$^2$/ms)&$\langle v\rangle$\,(nK/ms) &$\tau_\mathrm{ini}$\,(ms)&$\tau_\mathrm{eq}$\,(ms)\\
\noalign{\smallskip}\hline\noalign{\smallskip}
140 &0.08&$ -0.00246$&130&600 \\ 
280 &0.16&$-0.00492$&65&300\\
400&0.229&$-0.00705$&46&210\\
800 &0.457 &$-0.01406$&23&105 \\
\noalign{\smallskip}\hline
\end{tabular}
\end{center}
\end{table}
We use this calculation of the constant transport coefficients to later normalize the energy-dependent transport coefficients.
% Normalizability of the distribution functions requires the transport coefficients to converge towards zero for $\epsilon \rightarrow \infty$.
 To model energy-dependent transport coefficients that obey Eq.\,(\ref{eq:25}), have an energy-average matching the value of the constant transport coefficients, and constant or zero limits for large energies, a wide class of smooth functions is available. To accelerate thermalization in the region $\epsilon > \epsilon_\text i$, the diffusion in this energy region must be increased compared to the constant-coefficient case. Since the thermal tail does not contribute significantly to the dynamics of condensate formation and the number of particles in the thermal cloud for very large energies $\epsilon \gg \epsilon_\text i$ is small, we set the limit of the transport coefficients for $\epsilon\rightarrow\infty$ to zero. 
These conditions are fulfilled when using a Boltzman-like ansatz for the diffusion coefficient
\begin{equation}\label{eq:28}
    D(\epsilon) = \alpha \epsilon e^{-\beta \epsilon} \, ,
\end{equation}
where $\alpha$ and $\beta$ are constants yielding a normalized function such that the mean value $\langle D(\epsilon) \rangle = D$ matches the constant values calculated from Eqs.\,(\ref{eq:26}) and (\ref{eq:27}). This yields for $\beta = 1/T_\text f $
\begin{equation}
    \langle D(\epsilon)\rangle = \frac{1}{L}\int_0^L D(\epsilon)\, d\epsilon = \frac{\alpha T_\text f (T_\text f - e^{-L/T_\text f}(L + T_\text f))}{L}\, .
\end{equation}
As an example, for a large enough domain of at least $L \gg 100\,\text{nK}$ and $T_\text f = 32.5\, \text{nK}$ as required from the Cambridge $^{39}$K data \cite{gli21}, the relation $\alpha = \langle D \rangle / 2.5$ is obtained. Together with $D/v = -T_\text f$ this yields energy-dependent diffusion and drift coefficients as shown in Fig.\,\ref{fig1}.
\begin{figure*}
%\resizebox{0.48\textwidth}{!}%
  \includegraphics[scale=1.1]{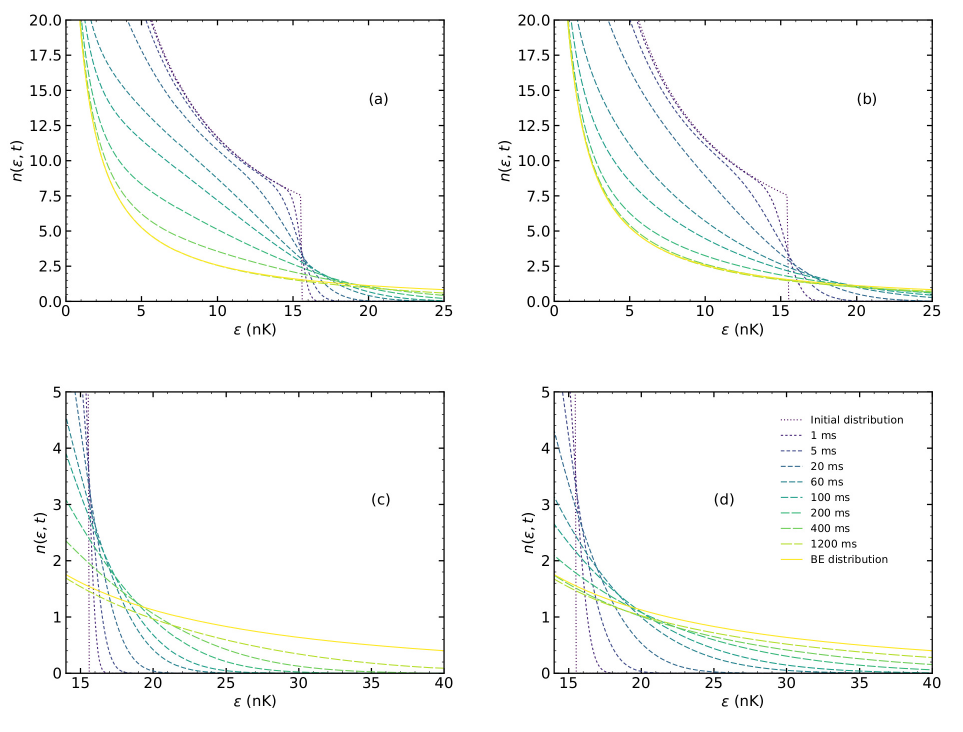}
\caption{Nonequilibrium evolution of quenched $^{39}$K  vapor calculated from the {exact} analytical solutions (a,c) of the nonlinear boson diffusion equation with constant transport coefficients and fixed chemical potential compared to the numerical solutions (b,d) of the NBDE with energy-dependent transport coefficients. The initial state is a Bose--Einstein distribution with $T_\text i=130\,\text{nK}$, $\mu_\text i=-0.67\,\text{nK}$ truncated at $\epsilon_\text i=15.56\,\text{nK}$, smallest dash length. The constant transport coefficients are $\langle D\rangle=0.08\,(\text{nK})^2/\text{ms}$ and $\langle v\rangle=-0.00246\,\text{nK}/\text{ms}$ {for $a=140\,a_0$}. The final temperature is $T_\text f=-D/v=32.5\,\text{nK}$, with the corresponding Bose--Einstein distribution (solid curves). The time evolution of the single-particle occupation number distribution is shown at $t=1,\,5,\,20,\,60,\,100,\,200,\,400\;\text{and}\;1200\,\text{ms}$ (increasing dash lengths{, same timesteps in all four frames}). {The analytical calculations for constant coefficients in (a), (c) show identical results, but (c) zooms into UV energies; accordingly for the numerical results with energy-dependent coefficients shown in in (b), (d).} }
\label{fig2}       % Give a unique label
\end{figure*}
\begin{figure}
  \includegraphics[scale=0.56]{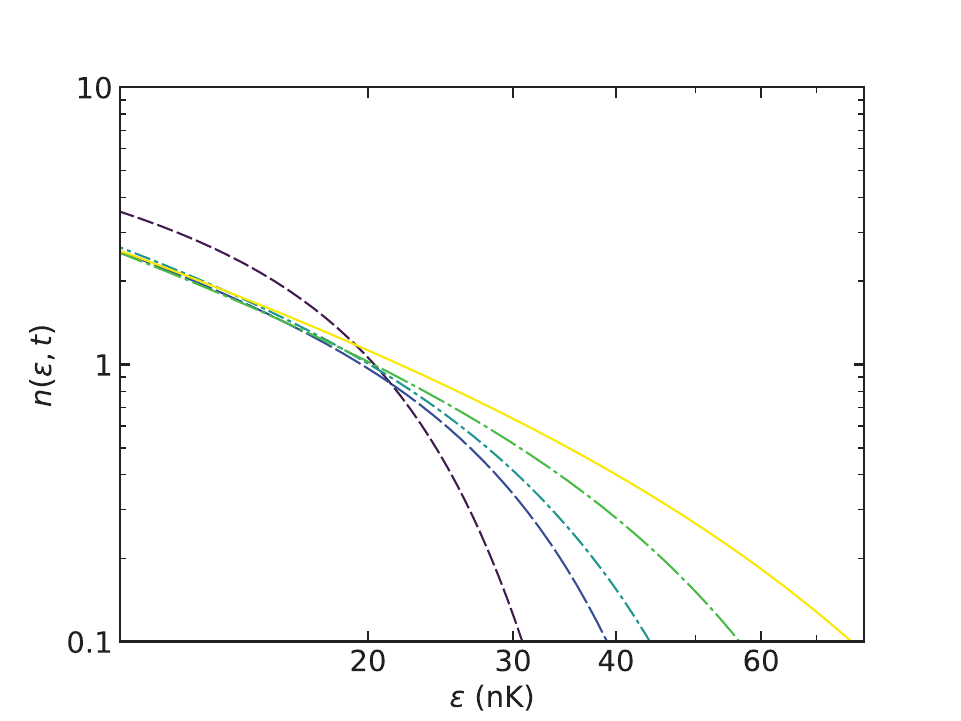}
\caption{ Nonequilibrium UV evolution of quenched $^{39}$K vapor calculated from the analytical and numerical solutions of the nonlinear boson diffusion equation for $\mu = 0$, using a double-log scale. The dashed lines are the exact analytical NBDE solutions with constant transport coefficients $D, v$ {and $a=140\,a_0$} for the timesteps $t=400$ and $1200\,\text{ms}$ (increasing dash lengths). The dot-dashed lines are the numerical solutions with energy-dependent transport coefficients $D(\epsilon)$ and $v(\epsilon)$ at the corresponding  timesteps. The solid curve is the Bose--Einstein distribution for the thermalized system. }
\label{fig3}       % Give a unique label
\end{figure}

%To assess the effects of the energy-dependent transport coefficients, one has to solve the nonlinear diffusion model numerically from scratch without relying on the analytical solution. %For some explicit energy-dependent transport coefficients a variety of nonlinear diffusion equations is analytically solvable by a nonlinear transformation \cite{Ivanova2007}. %However, for a general energy dependence and transport coefficients of the form of Eq.\,( \ref{eq:28}) a numerical treatment is essential. 
\subsection{Numerical solution of the nonlinear diffusion equation}
Contrary to the case of constant transport coefficients, the NBDE (\ref{nbde}) for energy-dependent transport coefficients is not known to be solvable via a nonlinear transformation, a numerical solution is required. The library for scientific machine learning of the programming language Julia offers a wide range of computationally strong algorithms to deal with nonlinear partial differential equations \cite{rackauckas2017differentialequations}. For Julia v1.6.7, the package MethodOfLines v0.10.0 was used for a finite difference discretization of a symbolically defined partial differential equation. The method of lines (MOL) discretization is a scheme in which all but one dimension are discretized \cite{Hamdi:2007}. The differential equation is reduced to a single continuous dimension, and hence the solution can be computed via methods of numerical integration of ordinary differential equations. The computations shown in Fig.\,\ref{fig2}\,(b,d) and Fig.\,\ref{fig3} are done for a domain $\epsilon \in [0, 200]$ and $t \in [0, 1200]$ with boundaries $n(\epsilon, t=0) = n_\text i$, $n(\epsilon = \mu, t) = \infty$ and $n(\epsilon = 200, t) = 0$, where $n_\text i$ is a quenched Bose--Einstein distribution.

Comparing the UV tail of the numerical solution with energy-dependent transport coefficients in Fig.\,\ref{fig2}\,(b,d) to the analytical solution \cite{gw22,kgw22} of the nonlinear diffusion model with constant transport coefficients in Fig.\,\ref{fig2}\,(a,c), or in the double-log-scale of Fig.\,\ref{fig3}, a significantly faster thermalization for energies $\epsilon > \epsilon_\text i$ occurs in the energy-dependent case, as required.
The temporal dynamics of the distribution function for energies smaller than the quench-energy $\epsilon_\text i$ does not  change significantly with energy-dependent coefficients. 
For times larger than $1200\,\text{ms}$ the system is assumed to be in equilibrium since it has reached twice the experimental thermalization value of $\tau_\text{eq}=600\,\text{ms}$. 

There appear, however, two numerical artifacts due to the fact that the calculations are done on a finite and discrete map. Since the value for infinity can only be covered by a very large but finite number, the boundary $n(\epsilon = \mu, t) = \infty$ is not equivalent to the pole of the analytical and equilibrium solution. Therefore, a small deviation of the numerical solution compared to the equilibrium solution occurs at $\epsilon\lesssim0.1\,\text{nK}$. Secondly, because of the limited energy domain and the fixed boundary $n(\epsilon = 200, t) = 0$, the numerical solution is forced to drop, while the equilibrium solution has a non-zero thermal tail over the whole energy domain. Hence, even at $1200\,\text{ms}$ there is still a small deviation between numerical and thermal tail. Nevertheless, using the numerical solution for the energy-dependent drift and diffusion yields an improved thermalization of the UV tail, and no overshoot when calculating the condensate fraction, see Sect.\,IV. 

 For constant coefficients, the numerical solution coincides with the exact analytical result when plotted together, apart from the two above-mentioned small IR and UV artifacts that would appear in any numerical solution. At 10 nK, exact and numerical solutions agree within four decimal places in the $^{39}$K case with the Julia solution. We had previously obtained corresponding agreements for $^{23}$Na with \texttt{MATLAB} \cite{sgw21} and $^{39}$K using C++ \cite{kgw22}.

\subsection{Time-dependent chemical potential}
\label{mu}
To determine the occupation of the condensed state $N_\text c(t)$, one requires particle-number conservation, which is a necessary condition for condensate formation to occur. Up to this point, we have assumed a fixed chemical potential $\mu_\text i$ in our model calculations. Setting the chemical potential to zero instead of $\mu_\text i$ will ensure a description of the system right when a condensate starts to form in overoccupied systems, without taking into account the initiation time $\tau_\text{ini}$ that is needed for the chemical potential to reach the point of the phase transition. In contrast to relativistic systems where particles can emerge from the available energy, cold bosonic gases strictly adhere to particle-number conservation. This conservation principle can be enforced by adjusting the chemical potential for each timestep until it reaches zero. 

The particle number of the thermal cloud for a distribution function $\tilde{n}(\epsilon,t)$ with a time-dependent chemical potential $\mu(t)$ is given by
\begin{equation}
    N_\text{th}(t) = \int_0^\infty g(\epsilon) \tilde{n}(\epsilon,t)\, d\epsilon \, ,
\end{equation} 
where the density of states $g(\epsilon)$ for a three-dimensional
Bose gas in a box is given by $g(\epsilon)=g_0\sqrt{\epsilon}$ with $g_0=(2m)^{3/2} V/(4\pi^2)$,
as obtained from the substitution of a summation over the quantum numbers of the associated states with an energy integration \cite{pitaevskii2016bose}. For a harmonic oscillator potential the dependence is $g(\epsilon)=g_0^{\text{HO}}\epsilon^2$.

At each timestep, $\mu(t)$ is determined to maintain particle number conservation, ensuring $N_\text i = N_\text{th}$, until $\mu$ reaches zero and condensation starts. Any further difference between $N_\text i$ and $N_\text{th}$ is then attributed to the condensate \cite{sgw21}, such that
\begin{equation}
    N_\text i - N_\text{th}(t) \equiv N_\text c(t)\,, \quad \quad \text{for}\; \mu = 0 \, .
\end{equation}
\begin{figure}
  \includegraphics[scale=0.56]{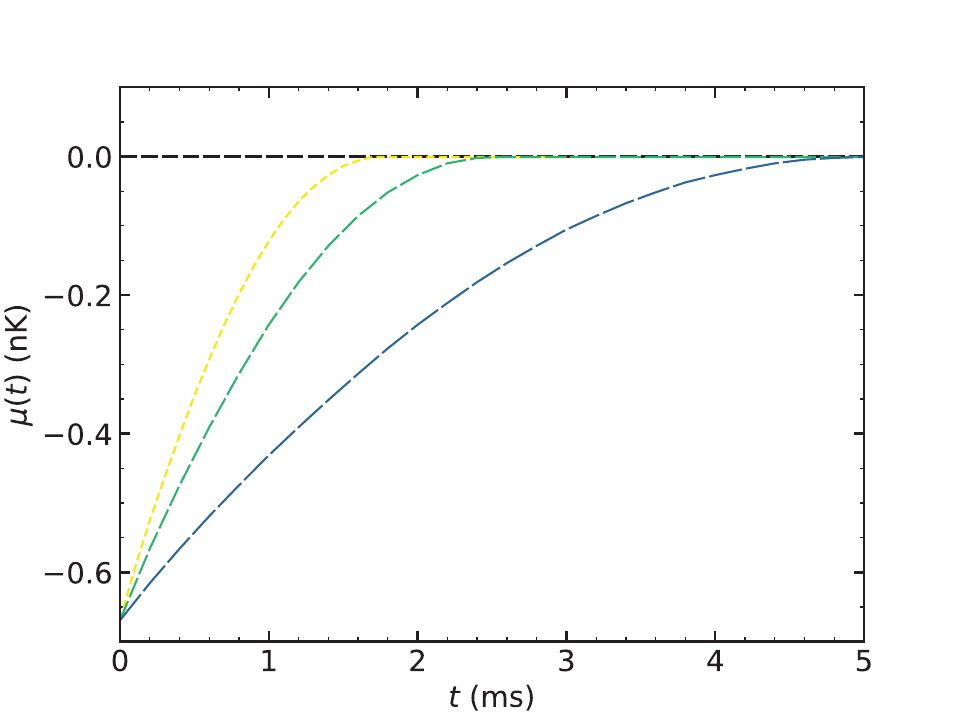}
\caption{ Chemical potential $\mu(t)$ following an energy-quench from $T_\text i=130\,\text{nK}$ to $T_\text f =32.5\,\text{nK}$ for  $^{39}$K vapor as calculated based on particle-number conservation from the num\,rical solution of the nonlinear bosonic diffusion equation with energy-dependent transport coefficients. The plot shows the time dependence for different scattering lengths $a = 140a_0, 280a_0$, and $400a_0$ (decreasing dash length). }
\label{fig4}       % Give a unique label
\end{figure}
The concept of a time-dependent chemical potential is, however, questionable, since the distribution function for $\mu_\text{i}\le \mu(t)<0$ is not in equilibrium. Hence, it does not come as a surprise that the condensate initiation time $\tau_\text{ini}$ for $^{39}$K vapor calculated  from the numerical solution with energy-dependent transport coefficients as shown in Fig.\,\ref{fig4} using particle-number conservation at each timestep comes out to be too small when compared to data. As an example, for $^{39}$K and a scattering length $a = 400\,a_0$, the calculation yields $\tau_\text{ini}\simeq 2$\,ms, which is significantly larger than a result obtained using constant transport coefficients, but still smaller than the experimental value of $46\,\text{ms}$ \cite{gli21}. The improvement is due to the fact that the thermalization of the tail of the distribution is faster and hence the constraint of particle-number conservation requires a slower adjustment of the chemical potential under the temporal dynamics of the nonlinear model. The large values for $\tau_\text{ini}$ found experimentally may be due to slow ergodic mixing of the trapped atomic cloud  \cite{kdg02} which is not explicitly taken into account in our model. For evaporative cooling in $^{87}$Rb at values of $\eta=\epsilon_\text{i}/T_\text{i}\geq 1.4$, measured initiation times \cite{kdg02} were found to agree with the results of a numerical model \cite{dgb00} based on quantum kinetic theory, but we are not aware of calculations for $\tau_\text{ini}$ in a deep-quench situation as in \cite{gli21} with $\eta=\epsilon_\text{i}/T_\text{i}= 0.12$. Hence, in the following we shall use the experimental values for 
$\tau_\text{ini}$ when comparing to data, keeping their scaling with $1/a$ as shown in Table\,\ref{tab1} and in Fig.\,\ref{fig5}. 

In our semiclassical description, the initiation time approaches zero for very large scattering length $a/a_0\rightarrow\infty$, whereas in a quantum model a small finite limit $\tau_0$ should be approached, as could probably be tested 
in experiments with small atoms such as $^7$Li at very large scattering lengths. The initiation time diverges for $a/a_0\rightarrow  0$, meaning that no condensate forms in a noninteracting system.

\begin{figure}
  \includegraphics[scale=0.33]{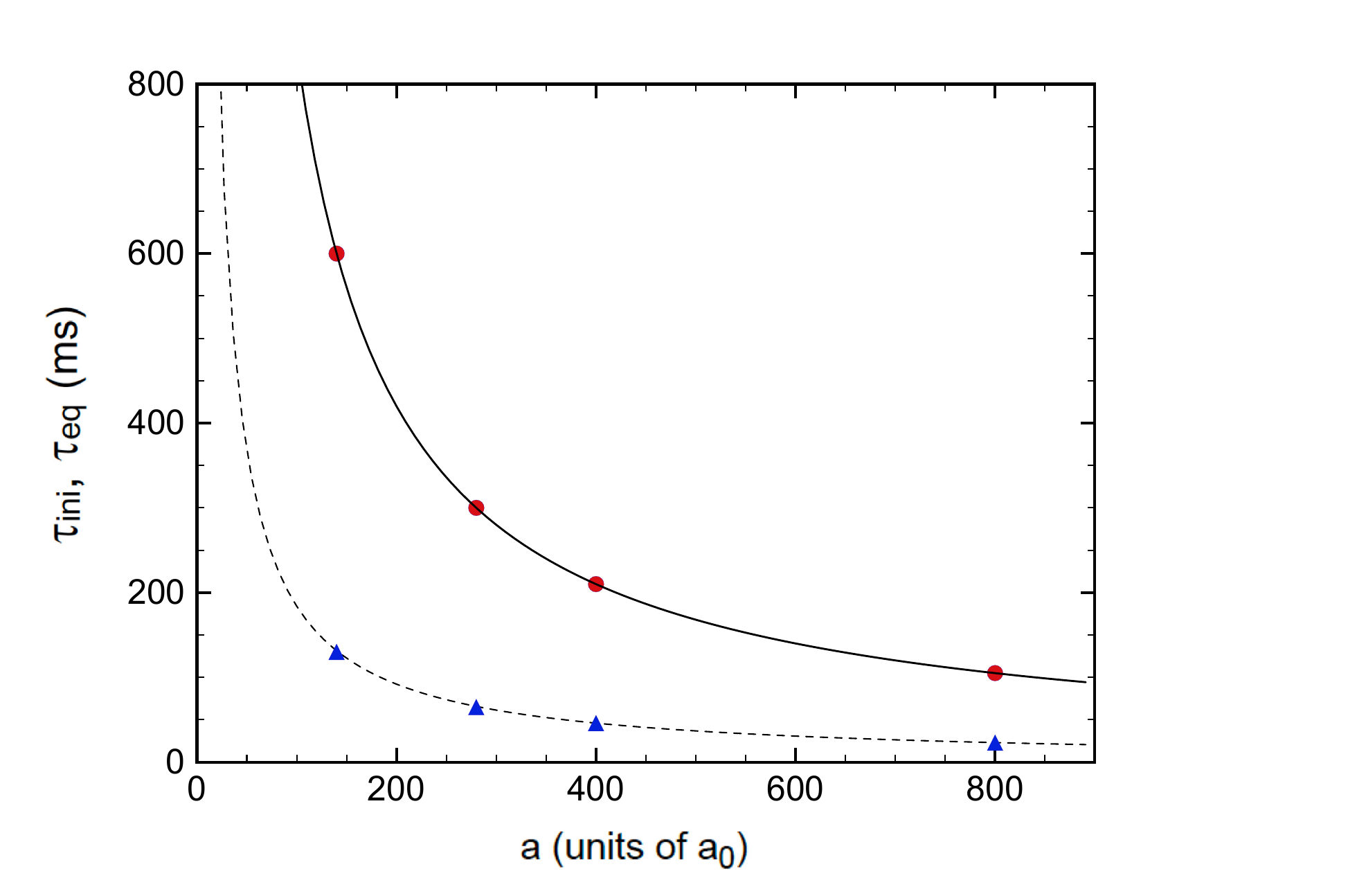}
\caption{Initiation time  $\tau_\text{ini}$ (dashed curve) and equilibration time $\tau_\text{eq}$ (solid, top) are proportional to the inverse scattering length. Values are according to time-dependent condensate formation data of $^{39}$K \cite{gli21}, see text. For the non-interacting limit $a \rightarrow 0$ the times diverge and the occupation of the condensed state takes infinitely long. }
\label{fig5}       % Give a unique label
\end{figure}
Regarding the scattering-length dependence of the equilibration times $\tau_\text{eq}$, in our previous investigation with constant transport coefficients for $^{39}$K \cite{kgw22} these were also taken to depend on $1/a$ in line with the experimentally determined scaling \cite{gli21}, and we keep this dependence in the present calculation for energy-dependent transport coefficients, see Fig.\,\ref{fig5}, upper solid curve. Whereas it may have been expected that the thermalization time depends on the inverse cross section \cite{sto91} -- and the transport coefficients on the cross section itself --, it has been shown that the dependence on the inverse interaction energies  \cite{kss92,svi91} is due to the emerging coherence between the highly occupied infrared modes, see Davis et al. \cite{dwg17} for a summary.

The equilibration time diverges for vanishing interaction energy as expected, but the limit  $\tau_\text{eq}\rightarrow 0$ for $a\rightarrow \infty$
it will have to be modified in a quantum description in order to remain larger than the initiation time. {Such extremely large scattering lengths are, however, not reached in the present investigation: In the region $140\le a/a_0 \le 800$, reasonable agreement with data} is achieved for the time-dependent condensate fraction, see next section.

\section{Time-dependent condensate formation} 

\begin{figure}
%\resizebox{0.49\textwidth}{!}{%
  \includegraphics[scale=0.58]{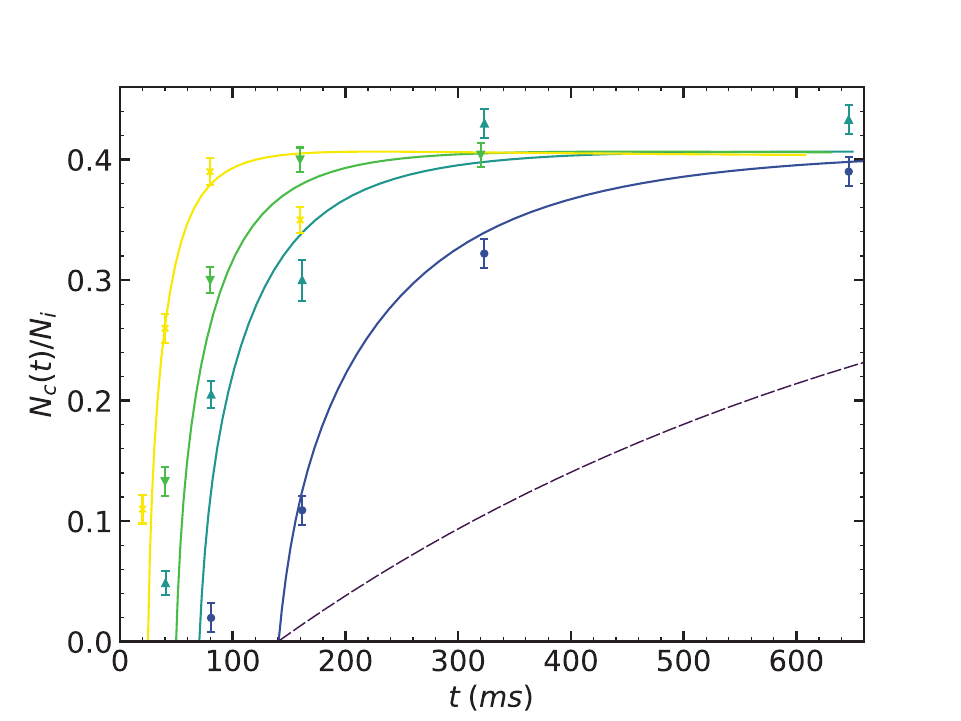}
\caption{Time-dependent condensate fraction for an equilibrating 3D Bose gas of $^{39}$K in an optical box for an energy quench at $\epsilon_\mathrm{i} = 15.56$ nK with $\mu_\mathrm{i} = -0.67$ nK, initial temperature $T_\mathrm{i} = 130$ nK and final temperature $T_\mathrm{f} = 32.5$ nK. The results were obtained from numerical solutions of the NBDE with energy-dependent transport coefficients {(solid curves)} according to the timescales and average transport coefficients from \cite{kgw22}. Results are shown for scattering lengths $a/a_0=$140 ({data displayed as} circles), 280 (triangles), 400 (inverted triangles) and 800 (crosses), plotted with {the respective} Cambridge data \cite{gli21}. The dashed line represents the result of the linear relaxation model for $a/a_0=$140.}
%$D=4.694$(nK)$^2$/ms and $v = -0.072$ nK/ms.
% The time axis has been shifted for best agreement with the data.}
\label{fig6}       % Give a unique label
\end{figure}

A numerical calculation of the time-dependent condensate fraction of $^{39}$K compared to the Cambridge data \cite{gli21}  is shown in Fig.\,\ref{fig6}. As $\langle D\rangle$ and          $\langle v\rangle$ depend on the equilibration times, the scattering lengths $a/a_0$, and the final temperature, respective pairs of energy-dependent transport coefficients were calculated with energy-averaged values according to Table\,\ref{tab1} that are taken from our Ref.\,\cite{kgw22} for constant coefficients. The time-dependent transition to the condensate is obtained with the additional condition of particle-number conservation at each timestep, as explained in the previous section. Once the parameters for the $a/a_0=140$ case are fixed, no further adjustment is done since the time scales are proportional to $1/a$ and the transport coefficients are inversely related to the equilibration time.

\begin{figure}
%\resizebox{0.49\textwidth}{!}{%
  \includegraphics[scale=0.58]{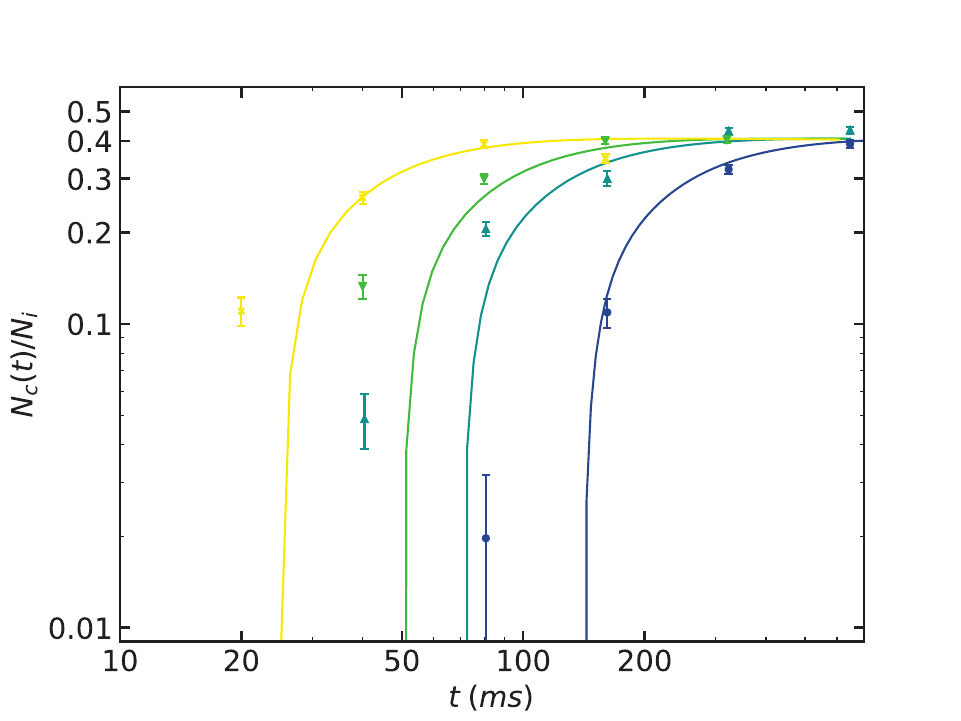}
\caption{Time-dependent condensate fraction of $^{39}$K as in Fig.\,\ref{fig6}, but using a double-log scale. The  numerical NBDE-solutions with energy-dependent transport coefficients are shown {as solid curves} for scattering lengths $a/a_0=$140 ({data displayed as} circles), 280 (triangles), 400 (inverted triangles) and 800 (crosses), plotted with Cambridge data \cite{gli21}. At small times, they obey power-law scaling. }
%The dashed line (with dot-dashed error limits) for $a/a_0=$140  at intermediate times represents a power-law with index 1.08(9) as suggested in Ref.\,\cite{gli21}.}
%$D=4.694$(nK)$^2$/ms and $v = -0.072$ nK/ms.
% The time axis has been shifted for best agreement with the data.}
\label{fig7}       % Give a unique label
\end{figure}

The NBDE-solutions with energy-dependent transport coefficients and the ensuing time-dependent (quasi-)~condensate fractions are compared with
%\footnote{\textit{quasi-condensate} fractions in Ref.\,\cite{gli21}} 
the data \cite{gli21} for various scattering lengths in Fig.\,\ref{fig6}. As in our previous work for constant coefficients \cite{kgw22}, we focus on $a/a_0=140, 280, 400$ and 800. The values of the transport coefficients and their energy dependence have not been further optimized with respect to the Cambridge data at the individual scattering lengths in view of the relatively sparse data points that are presently available. 

Condensation starts at the initiation time $\tau_\mathrm{ini}$, and rises towards the equilibrium value $N_\mathrm{c}^\mathrm{eq}/N_\mathrm{i}$. The rise is monotonic, no slight overshoot of the equilibrium value at large times occurs as may happen in the constant-coefficient case.
However, in particular for $a=140\,a_0$, deviations in the short-time behavior are observed,  because condensate formation in the model starts abruptly at the initiation time, whereas the $^{39}$K data may show a slow gradual increase, as had also been observed in the MIT $^{23}$Na data \cite{miesner1998}. In contrast, the $^{87}$Rb condensate-formation data with slow evaporative cooling \cite{kdg02} start at the initiation time and have very small error bars there. Should precise future deep-quench data show a gradual increase, one would have to introduce time-dependent transport coefficients.

As in our previous model with constant transport coefficients \cite{kgw22}, we confirm the experimental result \cite{gli21} that the time-dependent condensate fractions for different scattering lengths fall on a single universal curve when plotted as functions of $t'=ta/(300\,a_0)$, before reaching the equilibrium limit for $t'\rightarrow\infty$. 
Hence, the nonequilibrium system shows self-similar scaling \cite{gli21}, with a net flow of particles towards the infrared and into the condensed state. This is accompanied by a corresponding energy flow towards the ultraviolet carried by a small number of atoms that build up the Maxwell--Boltzmann tail, see  Figs.\,\ref{fig2} and \ref{fig3}.

%At all the scattering lengths investigated here, the onset of the time-dependent condensate fraction is consistent with a power law with an exponent close to one -- until about half %the equilibrium value is reached. This is revealed in a double-log plot of the condensate fractions that show almost linear rises until they bend over towards the equilibrium values, %consistent with the value $\alpha=1.15(8)$ for the power-law index found by the experimentalists from their data \cite{gli21}.

A double-log plot of our NBDE-results for energy-dependent transport coefficients in Fig.\,\ref{fig7} reveals that they obey power-law scaling only for very short times, whereas for intermediate times, they already bend over toward the equilibrium limit. 
%The dashed line with dot-dashed error limits indicates a power-law with index 1.08(9) as suggested in Ref.\,\cite{gli21}. 
Obviously, the index has to become time-dependent if one wants to use such laws to account for the time-dependent approach to equilibrium, as suggested in Ref.\,\cite{gli21}.

For a given initial distribution $n_\text i(\epsilon)$ such as the truncated equilibrium distribution in the present work, an approximate solution of the equilibration problem can also be obtained through the linear relaxation ansatz
that is sometimes used in descriptions of nonequilibrium problems, $\partial n_{\text{rel}} / \partial t = (n_{\text{eq}} - n_{\text{rel}})/\tau_{\text{eq}}$. It has the simple solution 
\begin{equation}
    n_{\text{rel}}(\epsilon,t) = n_\text i(\epsilon) e^{-t/\tau_{\text{eq}}} + n_{\text{eq}}(\epsilon)(1-e^{-t/\tau_{\text{eq}}})\, ,
\end{equation}
which enforces equilibration towards the thermal distribution $n_{\text{eq}}(\epsilon)$. When comparing the ensuing condensate fraction with the data, however, it turns
out (see Fig.\,\ref{fig6}, dashed curve) that the relaxation ansatz fails to account for time-dependent condensate formation due to the neglect of the nonlinearity of the process, which is closely modeled by the NBDE.

{The match between our results and the Cambridge data regarding the
time-dependent condensate fraction shown in Figs. 6,7 is certainly not perfect,
but it is way better than the outcome of a linear relaxation ansatz. The remaining discrepancy between our results and the data becomes particularly obvious 
in the double-log plot, Fig.\,\ref{fig7}. However, the general trend of the scattering-length-dependent
data is well reproduced, and we expect that the agreement will improve once data with better time resolution become available. These are indeed available for $^{87}$Rb \cite{kdg02}, but there for slow evaporative cooling, not for a deep quench that corresponds to the clean initial conditions that we are using here. Once deep-quench data with better time resolution become available, we shall aim
for a proper $\chi^2$-minimization of our results with respect to the data.}

Regarding the requirement of energy conservation in the system, we note that it is the total energy that must be conserved, referring to the time-dependent mean-field term
-- see Eq.\,(\ref{eq1}) -- plus the collision term and therefore, the collision term as expressed by the NDBE need not conserve the energy by itself. In contrast, it does conserve the particle number once the particles that move into the condensed state are also considered. Hence, we have used it here to compute the time-dependent condensate fraction.

\section{Summary and Conclusions}

We have modeled the thermalization and time-dependent condensate formation in a quenched Bose gas through solutions of the nonlinear boson diffusion equation.
Concentrating on $^{39}$K where excellent Cambridge data are available, we have introduced energy-dependent drift and diffusion coefficients. This requires a numerical solution of the NBDE, which is performed using the Julia programming language. In the limiting case of constant diffusion coefficient $D$ and drift coefficient $v$ that we had previously solved analytically with deep-quench initial conditions and appropriate boundary conditions at the singularity through a nonlinear transformation, the numerical results for the time-dependent distribution functions are found to agree with the exact analytical results apart from small deviations in the IR and the UV that are due to the finite integration domain of the numerical approach.

The time-dependent condensate fraction is obtained through the additional condition of particle-number conservation following the quench that -- in the particular $^{39}$K example -- removes 77\% of the atoms and ~97.5\% of the energy, reducing the temperature from $T_\text{i}=130$ nK to $T_\text{f}=32.5$ nK and eventually producing an equilibrium condensate fraction of $~ 40\%$. In the constant-coefficient case, the time-dependent approach of the condensate fraction to equilibrium may slightly overshoot the equilibrium value and later reach it from above, because a large constant drift towards the IR causes thermalization in the UV region to be too slow, such that at large times the Bose--Einstein limit is reached in the IR, but not yet in the UV. The resulting unbalanced energy-dependent equilibration has been remedied using energy-dependent drift and diffusion coefficients, requiring a numerical solution of the nonlinear diffusion equation.

In addition to particle-number conservation, the total energy must be conserved, referring to the time-dependent mean-field term
 plus the collision term. The collision term as expressed by the NDBE, however, need not conserve the energy by itself, although it does conserve the particle number once the particles that move into the condensed state are also considered. Hence, we have used it here to compute the time-dependent condensate fraction.

Agreement with the $^{39}$K data for the time-dependent  condensate fraction at various scattering lengths is achieved, confirming the experimental result that the condensate initiation time $\tau_\text{ini}$ and the equilibration time $\tau_\text{eq}$ scale with the inverse scattering length. The nonlinearity in the collision term as modeled through the NBDE solutions thus turns out to be appropriate for an understanding of the BEC formation data, whereas results of the linear relaxation-time ansatz severely undercut the data for the time-dependent condensate fraction.

{In the present cold-atom application of the nonlinear diffusion equation, we have thus improved the modeling through realistic energy-dependent transport coefficients instead of constant ones when comparing to the time-dependent condensate-formation data. There are other physical systems where the improvement is even more significant, so that the numerical solution method developed in this work will turn out to be useful.}

{This is particularly the case when the initial distribution is drastically different from the equilibrium one, as is the case for gluons in relativistic heavy-ion collisions: There, the initial nonequilibrium distribution of cold gluons is well approximated by a $\theta$-function up to the gluon saturation momentum; it is very different from the thermal Bose--Einstein distribution and the momentum dependence of the transport coefficients should not be neglected. In the cold-atom case, the initial nonequilibrium distribution is already quite similar to the final B--E distribution in the IR, such that the energy dependence of the transport coefficients is less relevant in the IR, although necessary to overcome the unphysical UV features of the constant-coefficient case.}
\section*{Acknowlegements}
ML thanks Johannes H\"olck for his help regarding the Julia code that has been developed for this work. We are grateful to Thomas Gasenzer for discussions and remarks.
%\end{acknowledgments}
%\bibliography{gw24_aps,larsson}
%apsrev4-2.bst 2019-01-14 (MD) hand-edited version of apsrev4-1.bst
%Control: key (0)
%Control: author (8) initials jnrlst
%Control: editor formatted (1) identically to author
%Control: production of article title (0) allowed
%Control: page (0) single
%Control: year (1) truncated
%Control: production of eprint (0) enabled
%

%\bibliographystyle{epj}
%\bibliography{gw22_epjd.bib}
%
\end{document}